\documentclass[ALICE,manyauthors]{cernphprep}
\pdfoutput=1

\usepackage[comma,square,numbers,sort&compress]{natbib}

\usepackage[bookmarks]{hyperref}
\usepackage[normalem]{ulem} 

\usepackage{color}
\definecolor{dgreen}{cmyk}{1.,0.,1.,0.2}        
\definecolor{orange}{cmyk}{0.,0.353,1.,0.}    

\newcommand{\mean}[1]{\langle #1 \rangle}
\newcommand{\be}{\begin{eqnarray}}
\newcommand{\ee}{\end{eqnarray}}

\def\snn{\mbox{$\sqrt{s_{_{\rm NN}}}$}}
\def\pb{Pb--Pb}
\def\pt{p_{\rm{T}}}

\def\psid{\Psi_{\rm 2}}

\def\gab{\langle \cos(\varphi_\alpha + \varphi_\beta - 2\psid) \rangle}
\def\dab{\langle \cos(\varphi_\alpha - \varphi_\beta) \rangle}
\def\gc{\gamma_{\alpha\beta}}
\def\dc{\delta_{\alpha\beta}}

\begin{document}

\begin{titlepage}
\PHyear{2017}
\PHnumber{241}      
\PHdate{12 September}  
%

\title{Constraining the magnitude of the Chiral Magnetic Effect with
Event Shape Engineering in \pb~collisions at $\mathbf{\sqrt{{\textit s}_{\rm NN}}}$= 2.76~TeV}
\ShortTitle{Constraining the Chiral Magnetic Effect at the LHC} 

\Collaboration{ALICE Collaboration}
\ShortAuthor{ALICE Collaboration} 

\begin{abstract}

In ultrarelativistic heavy-ion collisions, the event-by-event variation of the elliptic flow $v_2$ reflects 
fluctuations in the shape of the initial state of the system. This allows to select events with the same centrality but different initial 
geometry. This selection technique, Event Shape Engineering, has been used in the 
analysis of charge-dependent two- and three-particle correlations in \pb~collisions at $\snn =2.76$~TeV. The 
two-particle correlator $\dab$, calculated for different combinations of charges $\alpha$ and $\beta$, is almost 
independent of $v_2$ (for a given centrality), while the three-particle correlator $\gab$ scales almost linearly both 
with the event $v_2$ and charged-particle pseudorapidity density. The charge dependence of the three-particle 
correlator is often interpreted as evidence for the Chiral Magnetic Effect (CME), a parity violating effect of the strong 
interaction. However, its measured dependence on $v_2$ points to a large non-CME contribution to the correlator. Comparing the 
results with Monte Carlo calculations including a magnetic field due to the spectators, the upper limit of the CME signal contribution 
to the three-particle correlator in the 10--50\% centrality interval is found to be 26--33\% at 95\% confidence level.

\end{abstract}

\end{titlepage}
\setcounter{page}{2}

Parity symmetry is conserved in electromagnetism and is maximally violated in weak interactions. In 
strong interactions, global parity violation is not observed even though it is allowed by quantum chromodynamics. Local 
parity violation in strong interactions might occur in microscopic domains under conditions of finite 
temperature~\cite{Lee:1973iz, Lee:1974ma, Morley:1983wr, Kharzeev:1998kz} due to the existence of the topologically non-trivial configurations 
of the gluonic field, instantons and sphalerons. The interactions between 
quarks and gluonic fields with non-zero topological charge~\cite{Chern:1974ft} change the quark 
chirality. A local imbalance of chirality, coupled with the strong magnetic field produced in heavy-ion collisions 
($B \sim 10^{15}$~T)~\cite{Bzdak:2011yy, Deng:2012pc, Gursoy:2014aka}, would lead to charge separation 
along the direction of the magnetic field, which is on average perpendicular to the reaction plane (the plane of 
symmetry defined by the impact parameter vector and the beam direction), a phenomenon called Chiral Magnetic 
Effect (CME)~\cite{Kharzeev:2004ey, Kharzeev:2007tn, Kharzeev:2007jp, Fukushima:2008xe}. Since the sign of 
the topological charge is equally probable to be positive or negative, the charge separation averaged over many 
events is zero. This makes the observation of the CME experimentally difficult and possible only via 
correlation techniques.

Azimuthal anisotropies in particle production relative to the reaction plane, often referred to as anisotropic flow, 
are an important observable to study the system created in heavy-ion 
collisions~\cite{Voloshin:2008dg, Heinz:2013th}. Anisotropic flow arises from the asymmetry in the initial geometry 
of the collision. Its magnitude is quantified via the coefficients $v_n$ in a Fourier decomposition of the charged 
particle azimuthal distribution~\cite{Voloshin:1994mz,Poskanzer:1998yz}. Local parity violation would result in 
an additional sine term~\cite{Voloshin:2004vk}
\begin{equation}
\frac{{\rm d} N}{{\rm d}\Delta\varphi_{\alpha}} \sim 1 + 2v_{1,
  \alpha}\cos(\Delta\varphi_{\alpha}) + 2a_{1,
  \alpha}\sin(\Delta\varphi_{\alpha}) + 2v_{2,
  \alpha}\cos(2\Delta\varphi_{\alpha}) + ... ,
\end{equation}
where $\Delta\varphi_{\alpha} = \varphi_{\alpha} - \Psi_{\rm RP}$, $\varphi_{\alpha}$ is the azimuthal angle of the 
particle of charge $\alpha$ (+, $-$) and $\Psi_{\rm RP}$ is the reaction-plane angle. The first ($v_{1,\alpha}$) and 
the second ($v_{2,\alpha}$) coefficients are called directed and elliptic flow, respectively. The $a_{1, \alpha}$ coefficient
quantifies the effects from local parity violation. Since the average $\mean{a_{1,\alpha}}=0$ over many events, one can only 
measure $\mean{a_{1,\alpha}^2}$ or $\mean{a_{1,+} \, a_{1,-}}$. The charge-dependent two-particle correlator
\be
\dc \equiv \langle \cos(\varphi_{\alpha} - \varphi_{\beta}) \rangle  =
\langle \cos(\Delta\varphi_{\alpha}) \cos(\Delta\varphi_{\beta})
\rangle + \langle \sin(\Delta\varphi_{\alpha})
\sin(\Delta\varphi_{\beta}) \rangle
\ee
is not convenient for such a study, because along with the signal $\langle a_{1, \alpha} \, a_{1, \beta} \rangle$ ($\beta$ denotes the charge) there is 
a much stronger contribution from correlations unrelated to the azimuthal asymmetry in the initial geometry 
(``non-flow''). These correlations largely come from the inter-jet correlations and resonance decays. To increase the CME 
contribution it was proposed to use the following correlator~\cite{Voloshin:2004vk}
\be
\gc \equiv \langle \cos(\varphi_{\alpha} + \varphi_{\beta} - 2\Psi_{\rm RP})
\rangle = \langle \cos(\Delta\varphi_{\alpha})
\cos(\Delta\varphi_{\beta}) \rangle - \langle
\sin(\Delta\varphi_{\alpha}) \sin(\Delta\varphi_{\beta}) \rangle 
\ee
that measures the difference between the correlation projected onto the reaction plane and perpendicular 
to it. In practice, the reaction-plane angle is estimated by constructing the event plane angle 
$\psid$ using azimuthal particle distributions, which is why this correlator is often described as a 
three-particle correlator. This correlator suppresses background contributions at the level of $v_2$, the 
difference between the particle production in-plane and out-of-plane. Examples of such background sources are the
local charge conservation (LCC) coupled with elliptic flow~\cite{Schlichting:2010qia, Pratt:2010zn}, momentum 
conservation~\cite{Pratt:2010zn, Liao:2010nv, Bzdak:2010fd}, and directed-flow fluctuations~\cite{Teaney:2010vd}. The 
most significant background source for CME measurements is the LCC.

The measurements of 
charge-dependent azimuthal correlations performed at the Relativistic Heavy Ion 
Collider (RHIC)~\cite{Abelev:2009ac, Abelev:2009ad, Adamczyk:2013hsi, Adamczyk:2014mzf} and the 
Large Hadron Collider (LHC)~\cite{Abelev:2012pa, Adam:2015vje} are in qualitative agreement with the expectations 
for the CME. However, the interpretation of these experimental results is complicated due to possible 
background contributions. The 
Event Shape Engineering (ESE) technique was proposed to disentangle background contributions 
from the potential CME signal~\cite{Schukraft:2012ah}. This method makes it possible to select events with eccentricity
values significantly larger or smaller than the average in a given centrality class~\cite{Adam:2015eta, Aad:2015lwa} 
since $v_2$ scales approximately linearly with eccentricity~\cite{Gardim:2011xv}. Centrality estimates the degree of overlap between 
the two colliding nuclei, with low percentage values corresponding to head-on collisions. The CME contribution is expected to mainly scale 
with the magnetic field strength and to not have a strong dependence 
on the eccentricity~\cite{Bzdak:2011np}, while the background varies significantly. Therefore ESE provides a unique tool to separate the 
CME signal from the background for the three-particle correlator.

The CMS Collaboration has recently reported the measurement of the three-particle correlator $\gc$ in p--Pb collisions 
at $\sqrt{s_{_{\rm NN}}}=5.02$ TeV~\cite{Khachatryan:2016got}, where the direction of the magnetic field is expected to 
be uncorrelated to the reaction plane~\cite{Belmont:2016oqp}. The magnitude of the correlator in p--Pb and Pb--Pb collisions is 
comparable for similar final-state charged-particle multiplicities. This measurement indicates 
that the contribution of the CME to this observable in this multiplicity range is small.

In this paper we report the measurements of the two-particle correlator $\dc$, the three-particle correlator $\gc$, and the elliptic 
flow $v_2$ of unidentified charged particles. These measurements are performed for shape selected and unbiased events 
in \pb~collisions at $\snn = 2.76$~TeV. An upper limit on the CME contribution is deduced
from comparisons of the observed dependence of the correlations on the event $v_2$ to that estimated using 
Monte Carlo (MC) simulations of the magnetic field of spectators with different initial conditions. While this paper was in 
preparation, a paper employing a similar approach to estimate the fraction of the CME signal in the three-particle 
correlator was submitted by the CMS Collaboration~\cite{Sirunyan:2017quh}.

The data sample recorded by ALICE during the 2010 LHC \pb~run at $\snn= 2.76$~TeV is used for this analysis. 
General information on the ALICE detector and its performance can be found 
in~\cite{Aamodt:2008zz, Abelev:2014ffa}. The Time Projection Chamber (TPC)~\cite{Aamodt:2008zz, Alme:2010ke} 
and Inner Tracking System (ITS)~\cite{Aamodt:2008zz, Aamodt:2010aa} are used to reconstruct 
charged-particle tracks and measure their momenta with a track-momentum resolution better than 2\% for the transverse
momentum interval $0.2<\pt<5.0$~GeV/$c$~\cite{Abelev:2014ffa}. The two innermost layers of the ITS, the 
Silicon Pixel Detector (SPD), are employed for triggering and event selection. Two scintillator arrays 
(V0)~\cite{Aamodt:2008zz, Abbas:2013taa}, which cover the pseudorapidity 
ranges $-3.7<\eta<-1.7$ (V0C) and $2.8<\eta<5.1$ (V0A), are used for triggering, event selection, and the 
determination of centrality~\cite{Abelev:2013qoq} and $\Psi_{\rm 2}$. The trigger conditions and the event selection 
criteria are described in~\cite{Abelev:2014ffa}. An offline event selection is applied to remove beam induced 
background and pileup events. Approximately $9.8 \cdot 10^6$ minimum-bias \pb~events with a reconstructed 
primary vertex within $\pm 10$~cm from the nominal interaction point in the beam direction belonging to the 0--60\% centrality interval are used for this analysis.

Charged particles reconstructed using the combined information from the ITS and TPC in $|\eta|<0.8$ and 
$0.2<\pt<5.0$ GeV/$c$ are selected with full azimuthal coverage. Additional quality cuts are applied 
to reduce the contamination from secondary charged particles (i.e.\ particles originating from weak decays, conversions 
and secondary hadronic interactions in the detector material) and fake tracks (with random associations of space points). Only tracks 
with at least 70 space points in the TPC (out of a maximum of 159) with an average $\chi^2$ per degree-of-freedom for the track fit 
lower than 2, a distance of closest approach (DCA) to the reconstructed event vertex smaller than 2.4~cm in 
the transverse plane ($xy$) and 3.2~cm in the longitudinal direction ($z$) are accepted. The charged particle track reconstruction 
efficiency was estimated from HIJING simulations~\cite{Wang:1991hta, Gyulassy:1994ew} combined 
with a GEANT3~\cite{Brun:1994aa} detector model, and found to be independent of the collision centrality. The reconstruction efficiency 
of primary particles defined in~\cite{ALICE:2017prp}, which 
may bias the determination of the $\pt$ averaged charge-dependent correlations and flow, increases from 70\% at 
$\pt = 0.2$~GeV/$c$ to 85\% at $\pt \sim 1.5$~GeV/$c$ where it has a maximum. It then gradually decreases and is 
flat at 80\% for $\pt > 3.0$~GeV/$c$. The systematic uncertainty of the efficiency is about 5\%.

The event shape selection is performed as in~\cite{Adam:2015eta} based on the magnitude of the second-order reduced flow 
vector, $q_2$~\cite{Adler:2002pu}, defined as
\begin{equation}
q_2 = \frac{|\bf{Q}_{2}|}{\sqrt{M}}, 
\end{equation}
where $|{\bf Q}_{2}| = \sqrt{Q_{2, x}^2 + Q_{2, y}^2}$ is the magnitude of the second order harmonic flow 
vector and $M$ is the multiplicity. The vector ${\bf Q}_2$ is calculated from the azimuthal distribution of the energy 
deposition measured in the V0C. Its $x$ and $y$ components and the multiplicity are given by
\begin{equation}  
  Q_{2,x} = \sum_i w_i \cos(2 \varphi_i), \; Q_{2,y} = \sum_i w_i \sin(2 \varphi_i), \; M = \sum_i w_i,
\end{equation}
where the sum runs over all channels $i$ of the V0C detector ($i=1-32$), $\varphi_i$ is the azimuthal angle of channel 
$i$ and $w_i$ is the amplitude measured in channel $i$. The large gap in pseudorapidity ($|\Delta\eta|>0.9$) between the 
charged particles in the TPC used to determine $v_2$, $\dc$ and $\gc$ and those in the V0C suppresses 
non-flow effects. Ten event-shape classes with the lowest (highest) $q_2$ value corresponding to the 0--10\% (90--100\%) 
range are investigated for each centrality interval.

The flow coefficient $v_2$ is measured using the event plane method~\cite{Poskanzer:1998yz}. The orientation of the event 
plane $\psid$ is estimated from the azimuthal distribution of the energy deposition measured by the V0A detector. The 
event plane resolution is calculated from correlations between the event planes determined in the TPC and the two 
V0 detectors separately~\cite{Poskanzer:1998yz}. The non-flow contributions to the $v_2$ coefficient and charge-dependent 
azimuthal correlations are greatly suppressed by the large rapidity separation between the TPC and the V0A ($|\Delta\eta|>2.0$).

\begin{table}[tp] 
 \centering
  \begin{tabular}{ccc}
  \hline
    & Opposite charge & Same charge  \\
  \hline 
  $\dc$  & $(3.4 - 25) \times 10^{-5}$ &  $(3.1 - 10) \times 10^{-5}$ \\  
  $\gc$  & $(2.6 - 34) \times 10^{-6}$ &  $(4.1 - 74) \times 10^{-6}$ \\   
  \hline
  \hline
  $v_2$ & \multicolumn{2}{c} {$(1.2 - 4.7) \times 10^{-3}$ }\\ 
  \hline
  \end{tabular}
    \caption{Summary of absolute systematic uncertainties. The uncertainties depend on centrality and shape selection, whose minimum and maximum values are listed here.}
  \label{tab:syst}
\end{table}

\begin{figure}[t]
 \centering
 \includegraphics[keepaspectratio, width=0.65\columnwidth]{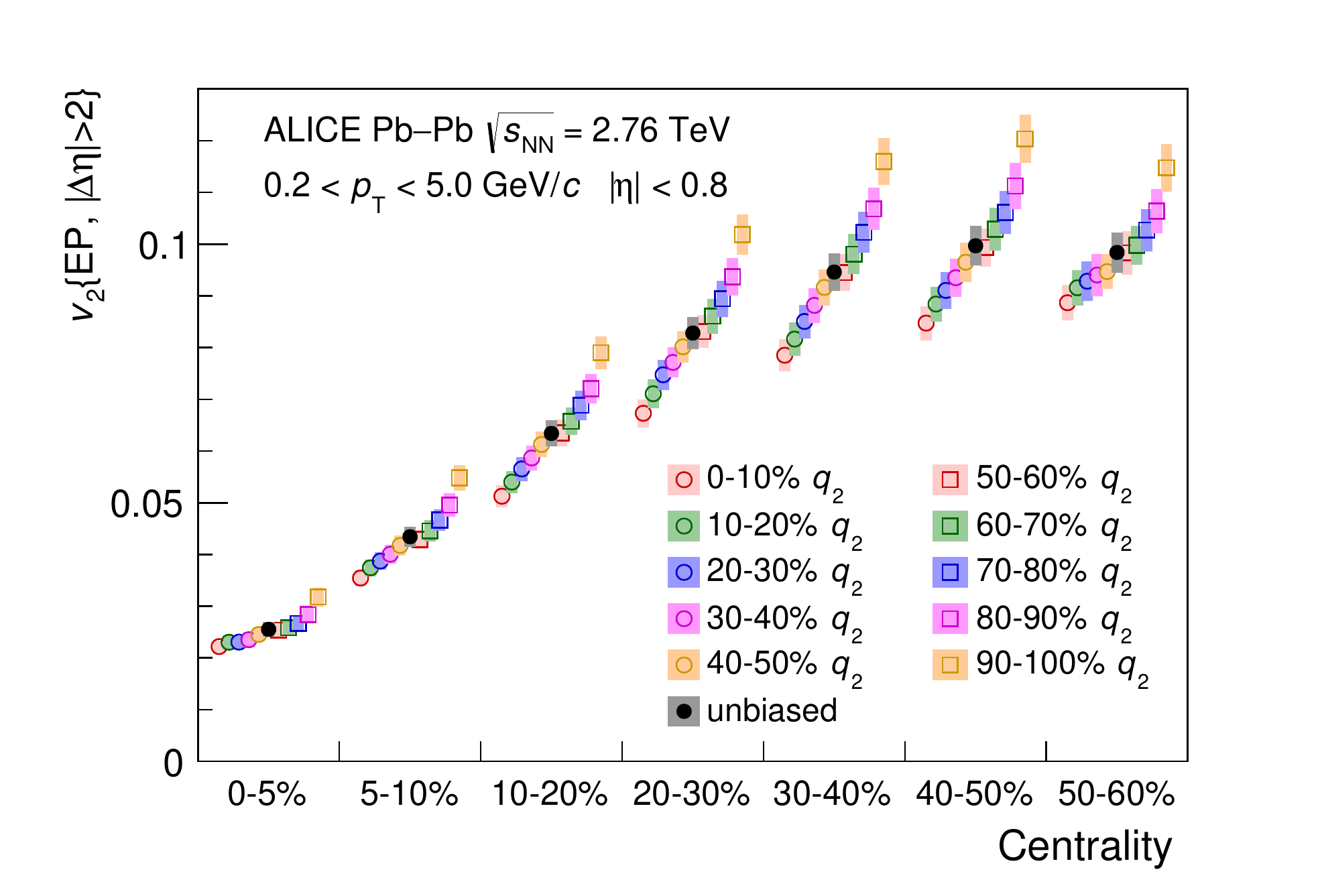}
 \caption{(Colour online) Unidentified charged particle $v_2$ for shape selected and unbiased events as a function of collision
   centrality. The event selection is based on $q_2$ determined in the V0C with the lowest (highest) value corresponding 
   to 0--10\% (90--100\%) $q_2$. Points are slightly shifted along the horizontal axis for better visibility. Error bars (shaded boxes) represent 
   the statistical (systematic) uncertainties.}
 \label{fig:vn_cent}
\end{figure}
\begin{figure}[t]
 \centering
 \includegraphics[keepaspectratio, width=0.65\columnwidth]{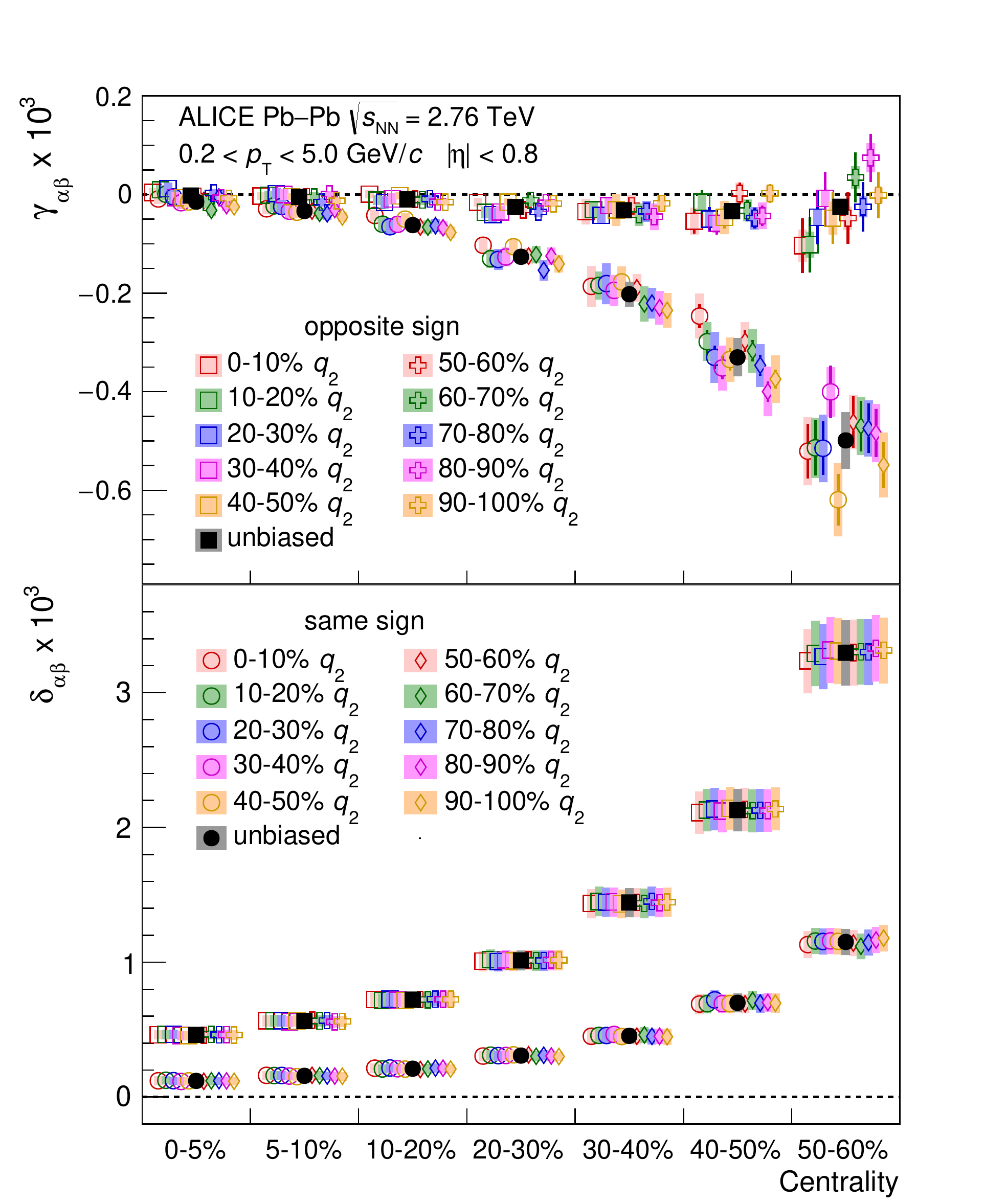}
 \caption{(Colour online) Top: Centrality dependence of $\gc$ for pairs of particles with same and opposite
   charge for shape selected and unbiased events. Bottom: Centrality dependence of $\dc$ for pairs of particles with same and opposite
   charge for shape selected and unbiased events.The event selection is based on $q_2$ determined in the V0C with the lowest 
   (highest) value corresponding to 0--10\% (90--100\%) $q_2$. Points are slightly
   shifted along the horizontal axis for better visibility in both
   panels. Error bars (shaded boxes) represent the statistical
   (systematic) uncertainties.}
 \label{fig:CME_cent}
\end{figure}

The absolute systematic uncertainties are evaluated from the variation of the results with different selection criteria on 
the reconstructed collision vertex, different magnetic field polarities, as well as by estimating the centrality from multiplicities 
measured by the TPC or the SPD rather than the V0 detector. Changes of the results
due to variations of the track-selection criteria (e.g.\ changing the DCA $xy$ and $z$ ranges, number of the TPC 
space points, using tracks reconstructed by the TPC only) are considered as part of the systematic 
uncertainties. The effect of reconstruction efficiency on the measurements is checked by randomly 
rejecting tracks to ensure a flat acceptance in $\pt$. The detector response is studied using 
HIJING and AMPT~\cite{Lin:2004en} simulations, where the $v_2$ coefficients and the charge-dependent azimuthal 
correlations obtained directly from the models are compared with those from reconstructed tracks. The largest contribution to 
the systematic uncertainties is given by the detector response. The checks related to the reconstruction efficiency, magnetic field polarity 
and track-selection criteria also yield significant deviations from the nominal values for $v_2$, $\gc$ and $\dc$, respectively. The contributions 
from all sources are added in quadrature as an estimate of the total systematic uncertainty. The resulting systematic 
uncertainties are summarized in Table~\ref{tab:syst}.

Figure~\ref{fig:vn_cent} presents the unidentified charged particle $v_2$ averaged over 
$0.2<\pt<5.0$ GeV/$c$ for shape selected and unbiased samples as a function of collision centrality. The 
measured $v_2$ for the shape selected events differs from the average by up to 25\%, which demonstrates that events 
with the desired initial spatial anisotropy can be experimentally selected. Sensitivity of the event shape selection deteriorates 
for peripheral collisions (already visible for the 50--60\% centrality class) due to the low multiplicity and for central collisions due to the reduced magnitude of flow~\cite{Adam:2015eta}.
\begin{figure}[t]
 \centering
 \includegraphics[keepaspectratio, width=0.65\columnwidth]{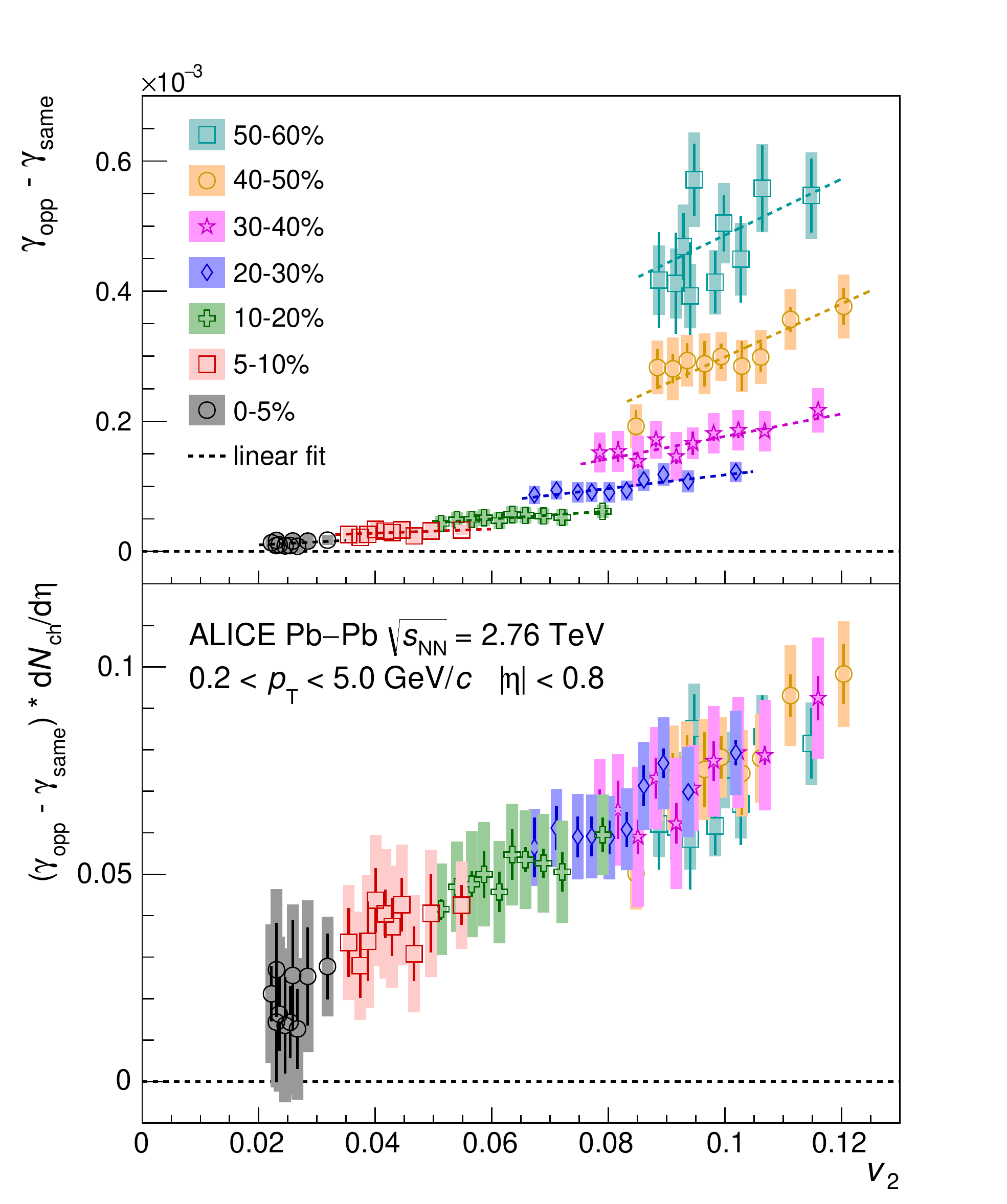}
 \caption{(Colour online) Top: Difference between opposite and same
   charge pair correlations for $\gc$
   as a function of $v_2$ for shape selected events together with a
   linear fit (dashed lines) for various centrality classes. Bottom:
   Difference between opposite and same charge pair correlations
   for $\gc$ multiplied by the charged-particle 
   density~\cite{Aamodt:2010cz} as a function of $v_2$ for shape
   selected events for various centrality classes. The event selection is based on $q_2$ determined in the V0C with the lowest 
   (highest) value corresponding to 0--10\% (90--100\%) $q_2$. Error bars (shaded
   boxes) represent the statistical (systematic) uncertainties.}
\label{fig:CMEdif_v2}
\end{figure}

The centrality dependence of $\gc$ for pairs of particles with same and opposite 
charge for shape selected and unbiased events is shown in the top panel of Fig.~\ref{fig:CME_cent}. The 
same charge results denote the average between pairs of particles with only positive and only negative charges 
since the two combinations are found to be consistent within statistical uncertainties. The correlation of pairs with 
the same charge is stronger than the correlation for pairs of opposite charge for both shape selected and 
unbiased events. The ordering of the correlations of 
pairs with same and opposite charge indicates a charge separation with respect to the reaction plane. The 
magnitude of the same and opposite charge pair correlations depends weakly on the event-shape selection 
($q_2$, i.e.\ $v_2$) in a given centrality bin.

The bottom panel of Fig.~\ref{fig:CME_cent} shows the centrality dependence of $\dc$ for pairs of particles with 
same and opposite charge for shape selected and unbiased samples. As reported in~\cite{Abelev:2012pa}, the 
magnitude of the correlation for the same charge pairs is smaller than for the opposite charge combinations. This is 
in contrast to the CME expectation, indicating that background dominates the correlations. The same 
and opposite charge pair correlations are insensitive to the event-shape selection in a given centrality bin.

The difference between opposite and same charge pair correlations for $\gc$ can 
be used to study the charge separation effect. This difference is presented as a function of 
$v_2$ for various centrality classes in the top panel of Fig.~\ref{fig:CMEdif_v2}. The difference is positive 
for all centralities and its magnitude decreases for more central collisions and with decreasing $v_2$ (in a 
given centrality bin). At least two effects could be responsible for the centrality dependence: the reduction of 
the magnetic field with decreasing centrality and the dilution of the correlation due to the increase in the 
number of particles~\cite{Abelev:2009ad} in more central collisions. The difference between opposite and same 
charge pair correlations multiplied by the charged-particle density in a given centrality bin, ${\rm d}N_{\rm ch}/{\rm d}\eta$ (taken 
from~\cite{Aamodt:2010cz}), to compensate for the dilution effect, is presented as a function of $v_2$ in the 
bottom panel of Fig.~\ref{fig:CMEdif_v2}. All the data points fall approximately onto the same line. This is 
qualitatively consistent with expectations from LCC where an increase in $v_2$, which modulates the correlation 
between balancing charges with respect to the reaction plane~\cite{Hori:2012kp}, results in a strong
effect. Therefore, the observed dependence on $v_2$ points to a large background contribution to $\gc$.
\begin{figure}[tp]
 \centering
 \includegraphics[keepaspectratio, width=0.65\columnwidth]{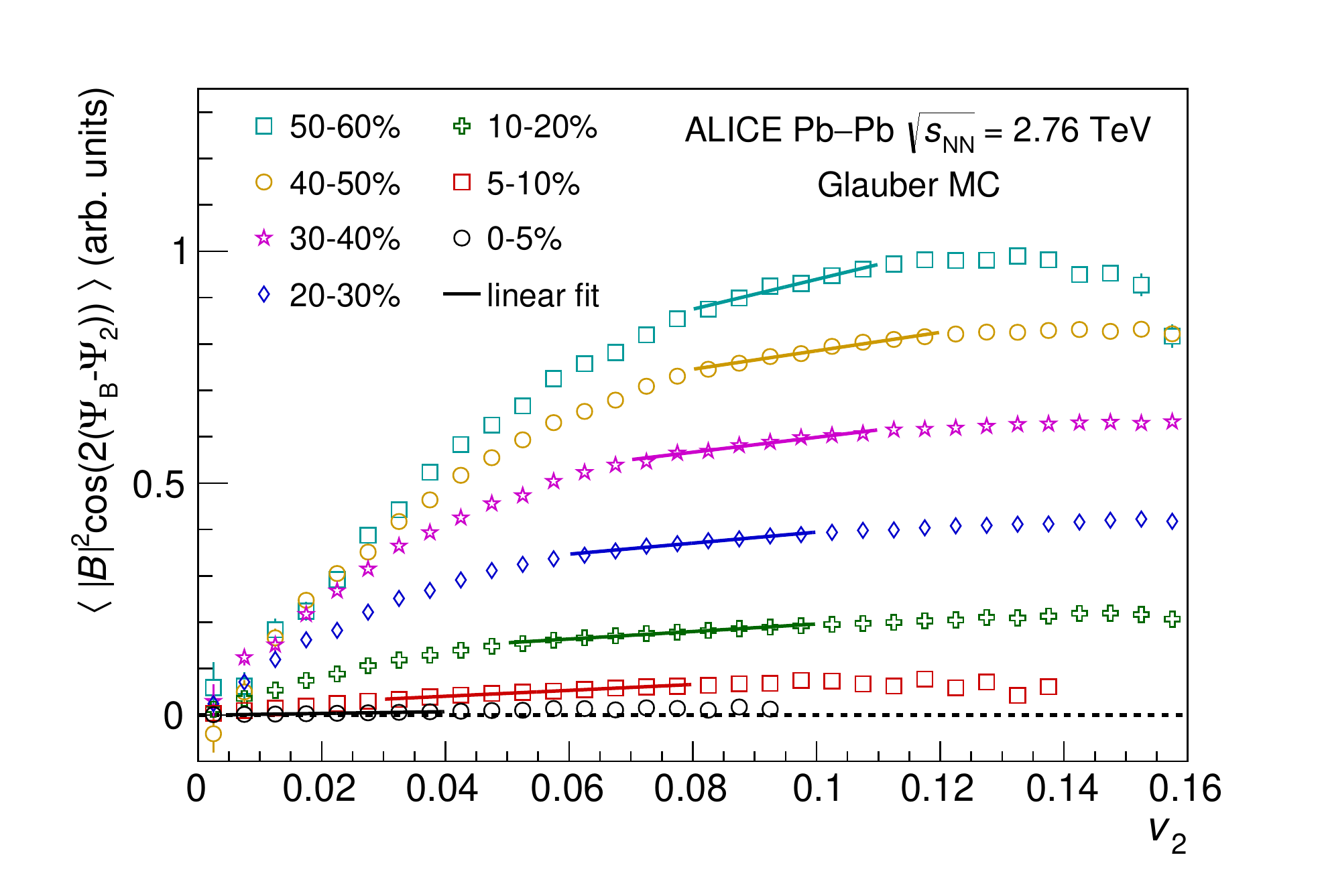}
 \caption{(Colour online) The expected dependence of the CME signal on $v_2$ for various centrality classes from 
 	a MC-Glauber simulation~\cite{Miller:2007ri} (see text for details). No event shape selection is performed 
	in the model, and therefore a large range in $v_2$ is covered. The solid lines depict linear fits based on the 
	$v_2$ variation observed within each centrality interval.}
 \label{fig:B2_Glb}
\end{figure}

The expected dependence of the CME signal on $v_2$ was evaluated with the help of a Monte Carlo Glauber~\cite{Miller:2007ri} 
calculation including a magnetic field. In this simulation, the centrality classes are determined from the 
multiplicity of charged particles in the acceptance of the V0 detector following the method presented in~\cite{Abelev:2013qoq}. The 
multiplicity is generated according to a negative binomial distribution with parameters taken 
from~\cite{Abelev:2013qoq} based on the number of participant nucleons and binary collisions. The 
elliptic flow is assumed to be proportional to the eccentricity of the participant nucleons and 
approximately reproduces the measured $\pt$-integrated $v_2$ values~\cite{Aamodt:2010pa}. The magnetic field is evaluated 
at the geometrical center of the overlap region from the number of spectator nucleons following 
Eq. (A.6) from~\cite{Kharzeev:2007jp} with the proper time $\tau=0.1$~fm/$c$. The magnetic field is calculated in 1\% 
centrality classes and averaged into the centrality intervals used for data analysis. It is assumed that the CME signal 
is proportional to $\langle |B|^2 \cos(2(\Psi_{\rm B} - \Psi_2)) \rangle$, where $|B|$ and $\Psi_{\rm B}$ are the magnitude and 
direction of the magnetic field, respectively. Figure~\ref{fig:B2_Glb} presents 
the expected dependence of the CME signal on $v_2$ for various centrality classes. Similar results are found using MC-KLN 
CGC~\cite{Drescher:2007ax, ALbacete:2010ad} and EKRT~\cite{Niemi:2015qia} initial conditions. The MC-KLN 
CGC simulation was performed using version 32 of the Monte Carlo $k_{\rm T}$-factorization code ($mckt$) 
available at~\cite{dumitru:CGCcode}, while the TRENTO model~\cite{Moreland:2014oya} was employed for EKRT 
initial conditions.
\begin{figure}[tp]
 \centering
 \includegraphics[keepaspectratio, width=0.65\columnwidth]{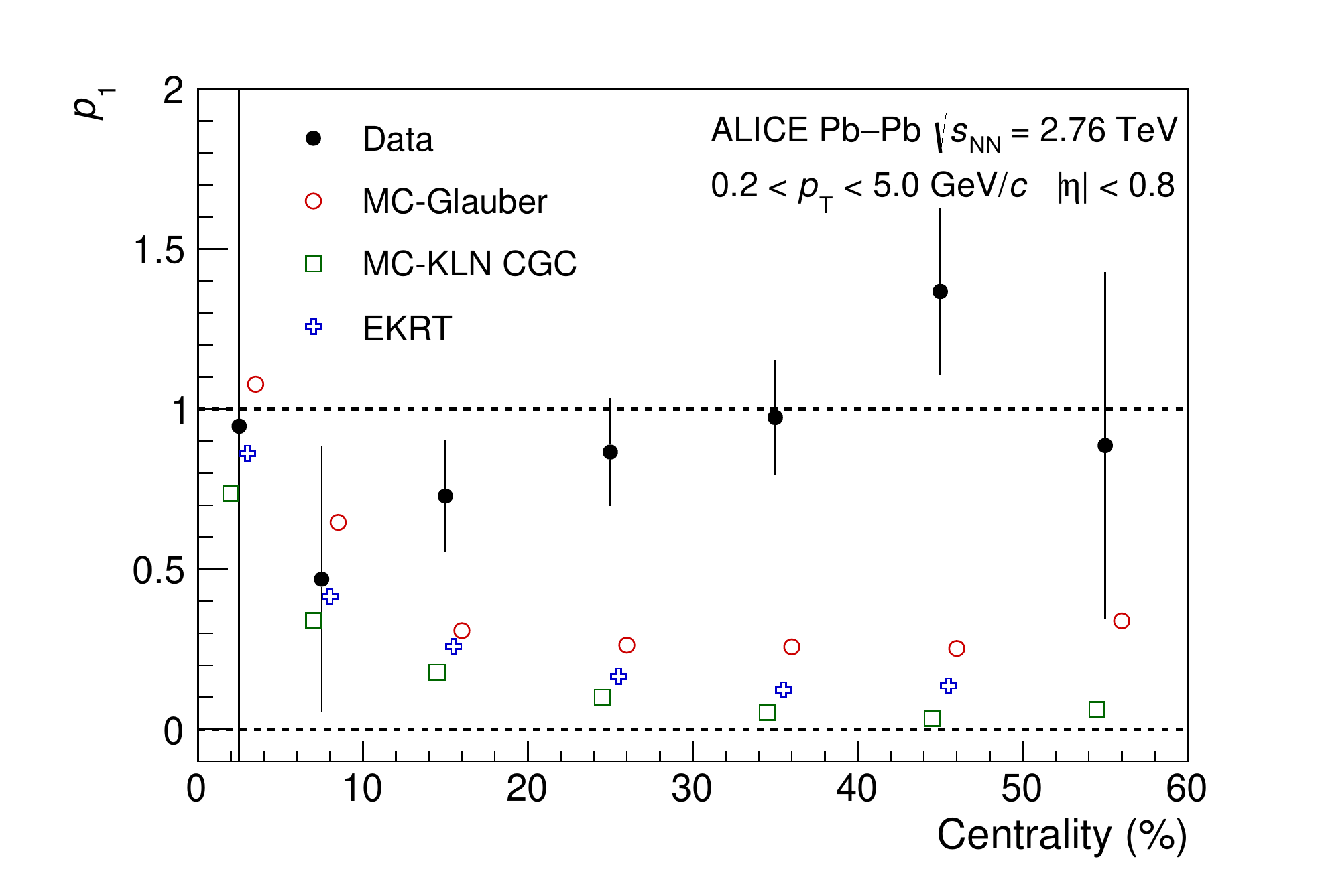}
 \caption{(Colour online) Centrality dependence of the $p_{1}$ parameter from
   a linear fit to the difference between opposite and same charge
   pair correlations for $\gc$ and from 
   linear fits to the CME signal expectations from MC-Glauber~\cite{Miller:2007ri}, MC-KLN
   CGC~\cite{Drescher:2007ax, ALbacete:2010ad} and
   EKRT~\cite{Niemi:2015qia} models (see text for details). Points
   from MC simulations are slightly shifted along the horizontal axis
   for better visibility. Only statistical uncertainties are shown.}
\label{fig:fit_slopes}
\end{figure}

To disentangle the potential CME signal from background, the dependence on $v_2$ of the difference between 
opposite and same charge pair correlations for $\gc$ and the CME signal expectations are fitted 
with a linear function (see lines in Figs.~\ref{fig:CMEdif_v2} (top panel) and~\ref{fig:B2_Glb}, respectively):
\be
F_1(v_2) = p_0(1 +  p_1(v_2 - \langle v_2 \rangle ) / \langle v_2 \rangle),
\ee
where $p_0$ accounts for the overall scale, which cannot be fixed in the MC calculations, 
and $p_1$ reflects the slope normalised such that in a pure background scenario, where the correlator
is directly proportional to $v_2$, it is equal to unity. The presence of a significant CME contribution, on the other hand, would result in non-zero 
intercepts at $v_2 = 0$ of the linear functions shown in Fig.~\ref{fig:CMEdif_v2}. The ranges used in these fits are based 
on the $v_2$ variation observed in data and the corresponding MC interval within each centrality range. The centrality dependence of 
$p_1$ from fits to data and to the signal expectations based on MC-Glauber, MC-KLN CGC and EKRT models is reported in 
Fig.~\ref{fig:fit_slopes}. The observed $p_1$ from data is a superposition of a possible CME signal and background. Assuming a 
pure background case, $p_1$ from data and MC models can be related according to
\be
f_{\rm CME} \times  p_{1,\rm{MC}} + (1 - f_{\rm CME}) \times 1 =  p_{1,\rm{data}},
\ee
where $f_{\rm CME}$ denotes the CME fraction to the charge dependence of $\gc$ and is given by 
\be
f_{\rm CME} = \frac{(\gamma_{\rm opp} - \gamma_{\rm same})^{\rm CME}} {(\gamma_{\rm opp} - \gamma_{\rm same})^{\rm CME} + (\gamma_{\rm opp} - \gamma_{\rm same})^{\rm Bkg}}.
\ee

\begin{figure}[tp]
 \centering
 \includegraphics[keepaspectratio, width=0.65\columnwidth]{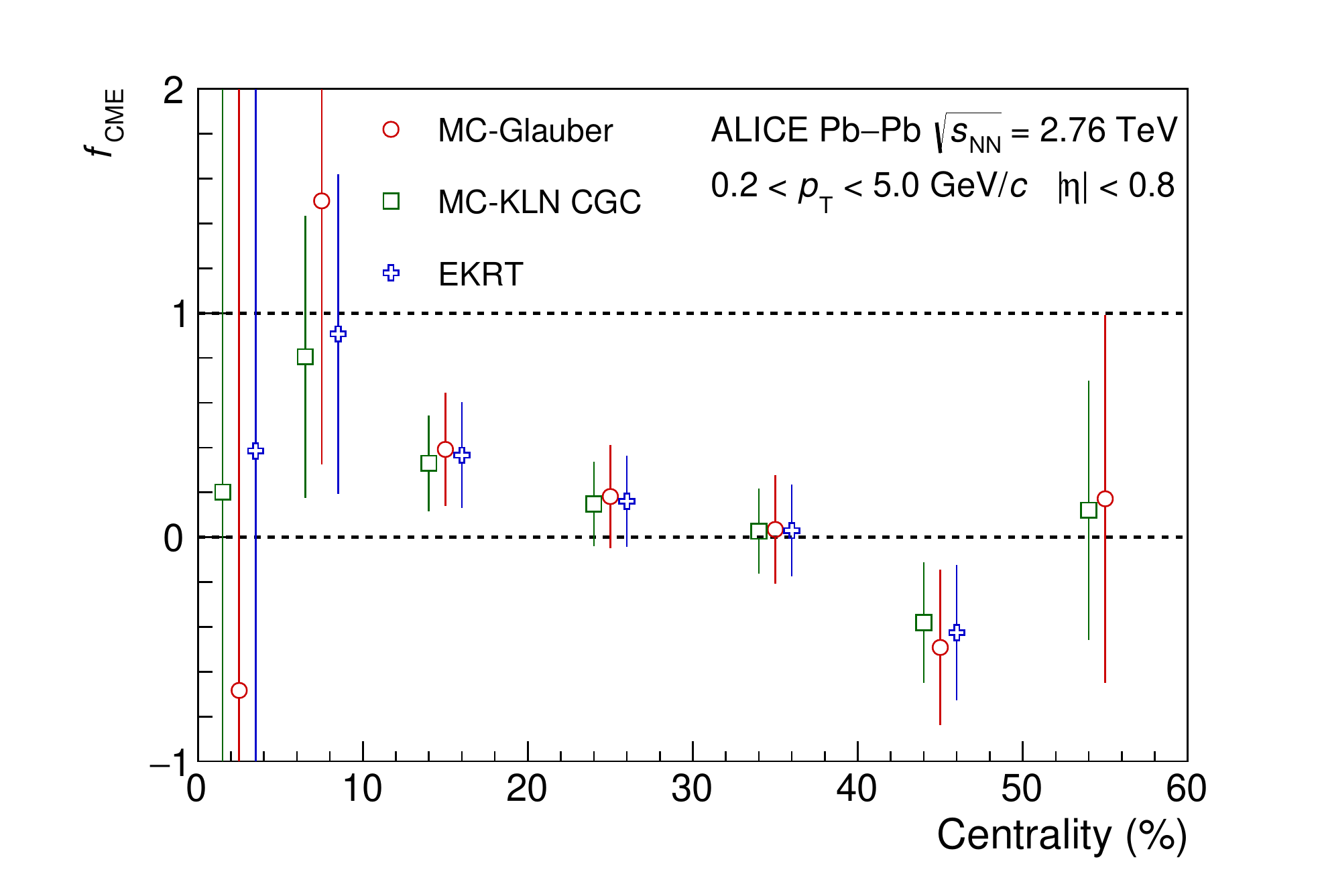}
 \caption{(Colour online) Centrality dependence of the CME fraction
   extracted from the slope parameter of fits to data and
   MC-Glauber~\cite{Miller:2007ri}, MC-KLN
   CGC~\cite{Drescher:2007ax, ALbacete:2010ad} and
   EKRT~\cite{Niemi:2015qia} models, respectively (see text for details). The dashed lines indicate the physical parameter space of the CME fraction. Points are slightly shifted along
   the horizontal axis for better visibility. Only statistical uncertainties are
   shown.}
\label{fig:cme_fraction}
\end{figure}

Figure~\ref{fig:cme_fraction} presents $f_{\rm CME}$ for the three models used in this study. The CME fraction 
cannot be precisely extracted for central (0--10\%) and peripheral (50--60\%) collisions due to the large statistical 
uncertainties on $p_1$ extracted from data. The negative values for the CME fraction obtained for the 40--50\% centrality range 
(deviating from zero by one~$\sigma$), if confirmed, would indicate that our expectations for the background 
contribution to be linearly proportional to $v_2$ are not accurate. Combining the points from 
10--50\% neglecting a possible centrality dependence gives $f_{\rm CME} = 0.10 \pm 0.13$, 
$f_{\rm CME} = 0.08 \pm 0.10$ and $f_{\rm CME} = 0.08 \pm 0.11$ for the MC-Glauber, MC-KLN CGC 
and EKRT models, respectively. These results are consistent with zero CME fraction and correspond to upper 
limits on $f_{\rm CME}$ of 33\%, 26\% and 29\%, respectively, at 95\% confidence level for the 10--50\% centrality interval. The CME fraction agrees with the observations in~\cite{Sirunyan:2017quh} where the centrality intervals overlap.

In summary, the Event Shape Engineering technique has been applied to measure the dependence on $v_2$ of the 
charge-dependent two- and three-particle correlators $\dc$ and $\gc$ in \pb~collisions at $\snn=2.76$~TeV. While for $\dc$
we observe no significant $v_2$ dependence in a given centrality bin, $\gc$ is found to be almost linearly dependent on $v_2$. When 
the charge dependence of $\gc$ is multiplied by the corresponding charged-particle density, to compensate for the dilution effect, a 
linear dependence on $v_2$ is observed consistently across all centrality classes. Using a Monte Carlo simulation with 
different initial-state models, we have found that the CME signal is expected to exhibit a weak dependence on $v_2$ in the 
measured range. The observations imply that the dominant contribution to $\gc$ is due to non-CME effects. In order to get a quantitative 
estimate of the signal and background contributions to the measurements, we fit both $\gc$ and the expected 
signal dependence on $v_2$ with a first order polynomial. This procedure allows to estimate the fraction of the CME signal in the centrality 
range 10--50\%, but not for the most central (0--10\%) and peripheral (50--60\%) collisions due to 
large statistical uncertainties. Averaging over the centrality range 10--50\% gives an upper limit of 26\% to 33\% (depending on 
the initial-state model) at 95\% confidence level for the CME contribution to the difference between opposite and 
same charge pair correlations for $\gc$.

\newenvironment{acknowledgement}{\relax}{\relax}
\begin{acknowledgement}
\section*{Acknowledgements}

The ALICE Collaboration would like to thank all its engineers and technicians for their invaluable contributions to the construction of the experiment and the CERN accelerator teams for the outstanding performance of the LHC complex.
The ALICE Collaboration gratefully acknowledges the resources and support provided by all Grid centres and the Worldwide LHC Computing Grid (WLCG) collaboration.
The ALICE Collaboration acknowledges the following funding agencies for their support in building and running the ALICE detector:
A. I. Alikhanyan National Science Laboratory (Yerevan Physics Institute) Foundation (ANSL), State Committee of Science and World Federation of Scientists (WFS), Armenia;
Austrian Academy of Sciences and Nationalstiftung f\"{u}r Forschung, Technologie und Entwicklung, Austria;
Ministry of Communications and High Technologies, National Nuclear Research Center, Azerbaijan;
Conselho Nacional de Desenvolvimento Cient\'{\i}fico e Tecnol\'{o}gico (CNPq), Universidade Federal do Rio Grande do Sul (UFRGS), Financiadora de Estudos e Projetos (Finep) and Funda\c{c}\~{a}o de Amparo \`{a} Pesquisa do Estado de S\~{a}o Paulo (FAPESP), Brazil;
Ministry of Science \& Technology of China (MSTC), National Natural Science Foundation of China (NSFC) and Ministry of Education of China (MOEC) , China;
Ministry of Science, Education and Sport and Croatian Science Foundation, Croatia;
Ministry of Education, Youth and Sports of the Czech Republic, Czech Republic;
The Danish Council for Independent Research | Natural Sciences, the Carlsberg Foundation and Danish National Research Foundation (DNRF), Denmark;
Helsinki Institute of Physics (HIP), Finland;
Commissariat \`{a} l'Energie Atomique (CEA) and Institut National de Physique Nucl\'{e}aire et de Physique des Particules (IN2P3) and Centre National de la Recherche Scientifique (CNRS), France;
Bundesministerium f\"{u}r Bildung, Wissenschaft, Forschung und Technologie (BMBF) and GSI Helmholtzzentrum f\"{u}r Schwerionenforschung GmbH, Germany;
General Secretariat for Research and Technology, Ministry of Education, Research and Religions, Greece;
National Research, Development and Innovation Office, Hungary;
Department of Atomic Energy Government of India (DAE) and Council of Scientific and Industrial Research (CSIR), New Delhi, India;
Indonesian Institute of Science, Indonesia;
Centro Fermi - Museo Storico della Fisica e Centro Studi e Ricerche Enrico Fermi and Istituto Nazionale di Fisica Nucleare (INFN), Italy;
Institute for Innovative Science and Technology , Nagasaki Institute of Applied Science (IIST), Japan Society for the Promotion of Science (JSPS) KAKENHI and Japanese Ministry of Education, Culture, Sports, Science and Technology (MEXT), Japan;
Consejo Nacional de Ciencia (CONACYT) y Tecnolog\'{i}a, through Fondo de Cooperaci\'{o}n Internacional en Ciencia y Tecnolog\'{i}a (FONCICYT) and Direcci\'{o}n General de Asuntos del Personal Academico (DGAPA), Mexico;
Nederlandse Organisatie voor Wetenschappelijk Onderzoek (NWO), Netherlands;
The Research Council of Norway, Norway;
Commission on Science and Technology for Sustainable Development in the South (COMSATS), Pakistan;
Pontificia Universidad Cat\'{o}lica del Per\'{u}, Peru;
Ministry of Science and Higher Education and National Science Centre, Poland;
Korea Institute of Science and Technology Information and National Research Foundation of Korea (NRF), Republic of Korea;
Ministry of Education and Scientific Research, Institute of Atomic Physics and Romanian National Agency for Science, Technology and Innovation, Romania;
Joint Institute for Nuclear Research (JINR), Ministry of Education and Science of the Russian Federation and National Research Centre Kurchatov Institute, Russia;
Ministry of Education, Science, Research and Sport of the Slovak Republic, Slovakia;
National Research Foundation of South Africa, South Africa;
Centro de Aplicaciones Tecnol\'{o}gicas y Desarrollo Nuclear (CEADEN), Cubaenerg\'{\i}a, Cuba, Ministerio de Ciencia e Innovacion and Centro de Investigaciones Energ\'{e}ticas, Medioambientales y Tecnol\'{o}gicas (CIEMAT), Spain;
Swedish Research Council (VR) and Knut \& Alice Wallenberg Foundation (KAW), Sweden;
European Organization for Nuclear Research, Switzerland;
National Science and Technology Development Agency (NSDTA), Suranaree University of Technology (SUT) and Office of the Higher Education Commission under NRU project of Thailand, Thailand;
Turkish Atomic Energy Agency (TAEK), Turkey;
National Academy of  Sciences of Ukraine, Ukraine;
Science and Technology Facilities Council (STFC), United Kingdom;
National Science Foundation of the United States of America (NSF) and United States Department of Energy, Office of Nuclear Physics (DOE NP), United States of America.    
\end{acknowledgement}

\bibliographystyle{utphys}   
\bibliography{biblio}

\newpage
\appendix
\section{The ALICE Collaboration}
\label{app:collab}

\begingroup
\small
\begin{flushleft}
S.~Acharya\Irefn{org139}\And 
J.~Adam\Irefn{org98}\And 
D.~Adamov\'{a}\Irefn{org95}\And 
J.~Adolfsson\Irefn{org34}\And 
M.M.~Aggarwal\Irefn{org100}\And 
G.~Aglieri Rinella\Irefn{org35}\And 
M.~Agnello\Irefn{org31}\And 
N.~Agrawal\Irefn{org48}\And 
Z.~Ahammed\Irefn{org139}\And 
N.~Ahmad\Irefn{org17}
% ...
\And
\renewcommand\labelenumi{\textsuperscript{\theenumi}~}

\section*{Affiliation notes}
\renewcommand\theenumi{\roman{enumi}}
\begin{Authlist}
\item \Adef{org*}Deceased
\item \Adef{orgI}Dipartimento DET del Politecnico di Torino, Turin, Italy
\item \Adef{orgII}Georgia State University, Atlanta, Georgia, United States
\item \Adef{orgIII}M.V. Lomonosov Moscow State University, D.V. Skobeltsyn Institute of Nuclear, Physics, Moscow, Russia
\item \Adef{orgIV}Department of Applied Physics, Aligarh Muslim University, Aligarh, India
\item \Adef{orgV}Institute of Theoretical Physics, University of Wroclaw, Poland
\end{Authlist}

\section*{Collaboration Institutes}
\renewcommand\theenumi{\arabic{enumi}~}
\begin{Authlist}
\item \Idef{org1}A.I. Alikhanyan National Science Laboratory (Yerevan Physics Institute) Foundation, Yerevan, Armenia
\item \Idef{org2}Benem\'{e}rita Universidad Aut\'{o}noma de Puebla, Puebla, Mexico
\item \Idef{org3}Bogolyubov Institute for Theoretical Physics, Kiev, Ukraine
\item \Idef{org4}Bose Institute, Department of Physics  and Centre for Astroparticle Physics and Space Science (CAPSS), Kolkata, India
\item \Idef{org5}Budker Institute for Nuclear Physics, Novosibirsk, Russia
\item \Idef{org6}California Polytechnic State University, San Luis Obispo, California, United States
\item \Idef{org7}Central China Normal University, Wuhan, China
\item \Idef{org8}Centre de Calcul de l'IN2P3, Villeurbanne, Lyon, France
\item \Idef{org9}Centro de Aplicaciones Tecnol\'{o}gicas y Desarrollo Nuclear (CEADEN), Havana, Cuba
\item \Idef{org10}Centro de Investigaciones Energ\'{e}ticas Medioambientales y Tecnol\'{o}gicas (CIEMAT), Madrid, Spain
\item \Idef{org11}Centro de Investigaci\'{o}n y de Estudios Avanzados (CINVESTAV), Mexico City and M\'{e}rida, Mexico
\item \Idef{org12}Centro Fermi - Museo Storico della Fisica e Centro Studi e Ricerche ``Enrico Fermi', Rome, Italy
\item \Idef{org13}Chicago State University, Chicago, Illinois, United States
\item \Idef{org14}China Institute of Atomic Energy, Beijing, China
\item \Idef{org15}COMSATS Institute of Information Technology (CIIT), Islamabad, Pakistan
\item \Idef{org16}Departamento de F\'{\i}sica de Part\'{\i}culas and IGFAE, Universidad de Santiago de Compostela, Santiago de Compostela, Spain
\item \Idef{org17}Department of Physics, Aligarh Muslim University, Aligarh, India
\item \Idef{org18}Department of Physics, Ohio State University, Columbus, Ohio, United States
\item \Idef{org19}Department of Physics, Pusan National University, Pusan, Republic of Korea
\item \Idef{org20}Department of Physics, Sejong University, Seoul, Republic of Korea
\item \Idef{org21}Department of Physics, University of Oslo, Oslo, Norway
\item \Idef{org22}Department of Physics and Technology, University of Bergen, Bergen, Norway
\item \Idef{org23}Dipartimento di Fisica dell'Universit\`{a} 'La Sapienza' and Sezione INFN, Rome, Italy
\item \Idef{org24}Dipartimento di Fisica dell'Universit\`{a} and Sezione INFN, Cagliari, Italy
\item \Idef{org25}Dipartimento di Fisica dell'Universit\`{a} and Sezione INFN, Trieste, Italy
\item \Idef{org26}Dipartimento di Fisica dell'Universit\`{a} and Sezione INFN, Turin, Italy
\item \Idef{org27}Dipartimento di Fisica e Astronomia dell'Universit\`{a} and Sezione INFN, Bologna, Italy
\item \Idef{org28}Dipartimento di Fisica e Astronomia dell'Universit\`{a} and Sezione INFN, Catania, Italy
\item \Idef{org29}Dipartimento di Fisica e Astronomia dell'Universit\`{a} and Sezione INFN, Padova, Italy
\item \Idef{org30}Dipartimento di Fisica `E.R.~Caianiello' dell'Universit\`{a} and Gruppo Collegato INFN, Salerno, Italy
\item \Idef{org31}Dipartimento DISAT del Politecnico and Sezione INFN, Turin, Italy
\item \Idef{org32}Dipartimento di Scienze e Innovazione Tecnologica dell'Universit\`{a} del Piemonte Orientale and INFN Sezione di Torino, Alessandria, Italy
\item \Idef{org33}Dipartimento Interateneo di Fisica `M.~Merlin' and Sezione INFN, Bari, Italy
\item \Idef{org34}Division of Experimental High Energy Physics, University of Lund, Lund, Sweden
\item \Idef{org35}European Organization for Nuclear Research (CERN), Geneva, Switzerland
\item \Idef{org36}Excellence Cluster Universe, Technische Universit\"{a}t M\"{u}nchen, Munich, Germany
\item \Idef{org37}Faculty of Engineering, Bergen University College, Bergen, Norway
\item \Idef{org38}Faculty of Mathematics, Physics and Informatics, Comenius University, Bratislava, Slovakia
\item \Idef{org39}Faculty of Nuclear Sciences and Physical Engineering, Czech Technical University in Prague, Prague, Czech Republic
\item \Idef{org40}Faculty of Science, P.J.~\v{S}af\'{a}rik University, Ko\v{s}ice, Slovakia
\item \Idef{org41}Faculty of Technology, Buskerud and Vestfold University College, Tonsberg, Norway
\item \Idef{org42}Frankfurt Institute for Advanced Studies, Johann Wolfgang Goethe-Universit\"{a}t Frankfurt, Frankfurt, Germany
\item \Idef{org43}Gangneung-Wonju National University, Gangneung, Republic of Korea
\item \Idef{org44}Gauhati University, Department of Physics, Guwahati, India
\item \Idef{org45}Helmholtz-Institut f\"{u}r Strahlen- und Kernphysik, Rheinische Friedrich-Wilhelms-Universit\"{a}t Bonn, Bonn, Germany
\item \Idef{org46}Helsinki Institute of Physics (HIP), Helsinki, Finland
\item \Idef{org47}Hiroshima University, Hiroshima, Japan
\item \Idef{org48}Indian Institute of Technology Bombay (IIT), Mumbai, India
\item \Idef{org49}Indian Institute of Technology Indore, Indore, India
\item \Idef{org50}Indonesian Institute of Sciences, Jakarta, Indonesia
\item \Idef{org51}INFN, Laboratori Nazionali di Frascati, Frascati, Italy
\item \Idef{org52}INFN, Laboratori Nazionali di Legnaro, Legnaro, Italy
\item \Idef{org53}INFN, Sezione di Bari, Bari, Italy
\item \Idef{org54}INFN, Sezione di Bologna, Bologna, Italy
\item \Idef{org55}INFN, Sezione di Cagliari, Cagliari, Italy
\item \Idef{org56}INFN, Sezione di Catania, Catania, Italy
\item \Idef{org57}INFN, Sezione di Padova, Padova, Italy
\item \Idef{org58}INFN, Sezione di Roma, Rome, Italy
\item \Idef{org59}INFN, Sezione di Torino, Turin, Italy
\item \Idef{org60}INFN, Sezione di Trieste, Trieste, Italy
\item \Idef{org61}Inha University, Incheon, Republic of Korea
\item \Idef{org62}Institut de Physique Nucl\'eaire d'Orsay (IPNO), Universit\'e Paris-Sud, CNRS-IN2P3, Orsay, France
\item \Idef{org63}Institute for Nuclear Research, Academy of Sciences, Moscow, Russia
\item \Idef{org64}Institute for Subatomic Physics of Utrecht University, Utrecht, Netherlands
\item \Idef{org65}Institute for Theoretical and Experimental Physics, Moscow, Russia
\item \Idef{org66}Institute of Experimental Physics, Slovak Academy of Sciences, Ko\v{s}ice, Slovakia
\item \Idef{org67}Institute of Physics, Academy of Sciences of the Czech Republic, Prague, Czech Republic
\item \Idef{org68}Institute of Physics, Bhubaneswar, India
\item \Idef{org69}Institute of Space Science (ISS), Bucharest, Romania
\item \Idef{org70}Institut f\"{u}r Informatik, Johann Wolfgang Goethe-Universit\"{a}t Frankfurt, Frankfurt, Germany
\item \Idef{org71}Institut f\"{u}r Kernphysik, Johann Wolfgang Goethe-Universit\"{a}t Frankfurt, Frankfurt, Germany
\item \Idef{org72}Institut f\"{u}r Kernphysik, Westf\"{a}lische Wilhelms-Universit\"{a}t M\"{u}nster, M\"{u}nster, Germany
\item \Idef{org73}Instituto de Ciencias Nucleares, Universidad Nacional Aut\'{o}noma de M\'{e}xico, Mexico City, Mexico
\item \Idef{org74}Instituto de F\'{i}sica, Universidade Federal do Rio Grande do Sul (UFRGS), Porto Alegre, Brazil
\item \Idef{org75}Instituto de F\'{\i}sica, Universidad Nacional Aut\'{o}noma de M\'{e}xico, Mexico City, Mexico
\item \Idef{org76}IRFU, CEA, Universit\'{e} Paris-Saclay, Saclay, France
\item \Idef{org77}iThemba LABS, National Research Foundation, Somerset West, South Africa
\item \Idef{org78}Joint Institute for Nuclear Research (JINR), Dubna, Russia
\item \Idef{org79}Konkuk University, Seoul, Republic of Korea
\item \Idef{org80}Korea Institute of Science and Technology Information, Daejeon, Republic of Korea
\item \Idef{org81}KTO Karatay University, Konya, Turkey
\item \Idef{org82}Laboratoire de Physique Subatomique et de Cosmologie, Universit\'{e} Grenoble-Alpes, CNRS-IN2P3, Grenoble, France
\item \Idef{org83}Lawrence Berkeley National Laboratory, Berkeley, California, United States
\item \Idef{org84}Moscow Engineering Physics Institute, Moscow, Russia
\item \Idef{org85}Nagasaki Institute of Applied Science, Nagasaki, Japan
\item \Idef{org86}National and Kapodistrian University of Athens, Physics Department, Athens, Greece
\item \Idef{org87}National Centre for Nuclear Studies, Warsaw, Poland
\item \Idef{org88}National Institute for Physics and Nuclear Engineering, Bucharest, Romania
\item \Idef{org89}National Institute of Science Education and Research, HBNI, Jatni, India
\item \Idef{org90}National Nuclear Research Center, Baku, Azerbaijan
\item \Idef{org91}National Research Centre Kurchatov Institute, Moscow, Russia
\item \Idef{org92}Niels Bohr Institute, University of Copenhagen, Copenhagen, Denmark
\item \Idef{org93}Nikhef, Nationaal instituut voor subatomaire fysica, Amsterdam, Netherlands
\item \Idef{org94}Nuclear Physics Group, STFC Daresbury Laboratory, Daresbury, United Kingdom
\item \Idef{org95}Nuclear Physics Institute, Academy of Sciences of the Czech Republic, \v{R}e\v{z} u Prahy, Czech Republic
\item \Idef{org96}Oak Ridge National Laboratory, Oak Ridge, Tennessee, United States
\item \Idef{org97}Petersburg Nuclear Physics Institute, Gatchina, Russia
\item \Idef{org98}Physics Department, Creighton University, Omaha, Nebraska, United States
\item \Idef{org99}Physics department, Faculty of science, University of Zagreb, Zagreb, Croatia
\item \Idef{org100}Physics Department, Panjab University, Chandigarh, India
\item \Idef{org101}Physics Department, University of Cape Town, Cape Town, South Africa
\item \Idef{org102}Physics Department, University of Jammu, Jammu, India
\item \Idef{org103}Physics Department, University of Rajasthan, Jaipur, India
\item \Idef{org104}Physikalisches Institut, Eberhard Karls Universit\"{a}t T\"{u}bingen, T\"{u}bingen, Germany
\item \Idef{org105}Physikalisches Institut, Ruprecht-Karls-Universit\"{a}t Heidelberg, Heidelberg, Germany
\item \Idef{org106}Physik Department, Technische Universit\"{a}t M\"{u}nchen, Munich, Germany
\item \Idef{org107}Purdue University, West Lafayette, Indiana, United States
\item \Idef{org108}Research Division and ExtreMe Matter Institute EMMI, GSI Helmholtzzentrum f\"ur Schwerionenforschung GmbH, Darmstadt, Germany
\item \Idef{org109}Rudjer Bo\v{s}kovi\'{c} Institute, Zagreb, Croatia
\item \Idef{org110}Russian Federal Nuclear Center (VNIIEF), Sarov, Russia
\item \Idef{org111}Saha Institute of Nuclear Physics, Kolkata, India
\item \Idef{org112}School of Physics and Astronomy, University of Birmingham, Birmingham, United Kingdom
\item \Idef{org113}Secci\'{o}n F\'{\i}sica, Departamento de Ciencias, Pontificia Universidad Cat\'{o}lica del Per\'{u}, Lima, Peru
\item \Idef{org114}SSC IHEP of NRC Kurchatov institute, Protvino, Russia
\item \Idef{org115}Stefan Meyer Institut f\"{u}r Subatomare Physik (SMI), Vienna, Austria
\item \Idef{org116}SUBATECH, IMT Atlantique, Universit\'{e} de Nantes, CNRS-IN2P3, Nantes, France
\item \Idef{org117}Suranaree University of Technology, Nakhon Ratchasima, Thailand
\item \Idef{org118}Technical University of Ko\v{s}ice, Ko\v{s}ice, Slovakia
\item \Idef{org119}Technical University of Split FESB, Split, Croatia
\item \Idef{org120}The Henryk Niewodniczanski Institute of Nuclear Physics, Polish Academy of Sciences, Cracow, Poland
\item \Idef{org121}The University of Texas at Austin, Physics Department, Austin, Texas, United States
\item \Idef{org122}Universidad Aut\'{o}noma de Sinaloa, Culiac\'{a}n, Mexico
\item \Idef{org123}Universidade de S\~{a}o Paulo (USP), S\~{a}o Paulo, Brazil
\item \Idef{org124}Universidade Estadual de Campinas (UNICAMP), Campinas, Brazil
\item \Idef{org125}Universidade Federal do ABC, Santo Andre, Brazil
\item \Idef{org126}University of Houston, Houston, Texas, United States
\item \Idef{org127}University of Jyv\"{a}skyl\"{a}, Jyv\"{a}skyl\"{a}, Finland
\item \Idef{org128}University of Liverpool, Liverpool, United Kingdom
\item \Idef{org129}University of Tennessee, Knoxville, Tennessee, United States
\item \Idef{org130}University of the Witwatersrand, Johannesburg, South Africa
\item \Idef{org131}University of Tokyo, Tokyo, Japan
\item \Idef{org132}University of Tsukuba, Tsukuba, Japan
\item \Idef{org133}Universit\'{e} Clermont Auvergne, CNRS/IN2P3, LPC, Clermont-Ferrand, France
\item \Idef{org134}Universit\'{e} de Lyon, Universit\'{e} Lyon 1, CNRS/IN2P3, IPN-Lyon, Villeurbanne, Lyon, France
\item \Idef{org135}Universit\'{e} de Strasbourg, CNRS, IPHC UMR 7178, F-67000 Strasbourg, France, Strasbourg, France
\item \Idef{org136}Universit\`{a} degli Studi di Pavia, Pavia, Italy
\item \Idef{org137}Universit\`{a} di Brescia, Brescia, Italy
\item \Idef{org138}V.~Fock Institute for Physics, St. Petersburg State University, St. Petersburg, Russia
\item \Idef{org139}Variable Energy Cyclotron Centre, Kolkata, India
\item \Idef{org140}Warsaw University of Technology, Warsaw, Poland
\item \Idef{org141}Wayne State University, Detroit, Michigan, United States
\item \Idef{org142}Wigner Research Centre for Physics, Hungarian Academy of Sciences, Budapest, Hungary
\item \Idef{org143}Yale University, New Haven, Connecticut, United States
\item \Idef{org144}Yonsei University, Seoul, Republic of Korea
\item \Idef{org145}Zentrum f\"{u}r Technologietransfer und Telekommunikation (ZTT), Fachhochschule Worms, Worms, Germany
\end{Authlist}
\endgroup
\end{document}